\documentstyle[multicol,aps]{revtex}
\begin{document}
\draft
\title{Stochastic Shell Model:
a model of anomalous scaling and non-Gaussian distribution in
turbulence}
\author{Hideki Takayasu\cite{taka}}
\address{Graduate School of Information Sciences, Tohoku University,
Aoba-ku, Sendai 980-77, Japan}
\author{Y-h. Taguchi\cite{tag}}
\address{Department of Physics, Tokyo Institute of Technology,
Oh-okayama,  Meguro-ku, Tokyo 152, Japan}
\author{Tomoo Katsuyama\cite{katsu}}
\address{Department of Physics, Tokyo Metropolitan University,
Minami-Ohsawa 1-1, Hachioji, Tokyo 192-03, Japan}
\date{\today}
\maketitle
\begin{abstract}
We propose a simple stochastic model of cascading
transport in wave number space to clarify the origin
of intermittent behavior of fully-developed fluid
turbulence. In spite of lack of nonlinearity and
 viscosity the model gives non-Gaussian
fluctuations and multifractal scalings consistent
with experimental data.
\end{abstract}
\pacs{47.27.Eq\hfill chao-dyn/9503004,TITCMT-95-7}
\begin{multicols}{2}
\narrowtext
Spatio-temporal structure of
fully-developed isotropic fluid turbulence\cite{Monin}
is one of frontier topics
in both fluid mechanics and statistical physics.
Although several new concepts, for example,
fractals\cite{fractal}, chaos\cite{chaos},
and intermittency, have been introduced to
characterize fully-developed turbulence,
our present understanding of fluid turbulence is
still far from complete.

There are many difficulties both in experimental
and numerical approaches.
Experimentally, typical available data
is time sequential data of velocity at a fixed point\cite{katsuyama},
which is obviously not sufficient to observe global spatio-temporal
structures of fluid turbulence.
Instantaneous local velocities at arbitrary points
can be obtained by using direct numerical integration\cite{kida},
however, fully-developed turbulence requires so many degrees
of freedom that even a super computer is not powerful enough.
To help our understanding of the behavior of turbulence we need
simple models which are easily accessible and share some basic
properties with fluid turbulence.

In the history of study on turbulence one of the most
successful approaches is Kolmogorov's scaling argument\cite{Monin},
in which the energy spectrum, $E(k)$, was predicted to
have a power law wave number dependence as $k^{-5/3}$.
His prediction was based on a very simple assumption of
cascading process that conserved energy is transported
continuously in the wave number space toward higher
wave numbers. Although his simple assumption captured
the most basic property of fully-developed turbulence,
a modification was proposed by himself and others to include
the effect of spontaneous spatial inhomogeneity
so called the intermittency\cite{batcheler}.
In order to take into account the
fluctuations of energy dissipation rates a log-normal distribution
was assumed and the exponent of the energy spectrum was
modified\cite{log-normal}.

Another approach to the intermittency is based on geometrical
models that an eddy breaks into smaller eddies forming a fractal
configuration\cite{Frisch}. This fractal approach is recently extended to
apply the concept of multifractals\cite{multi} and now the agreement with
experiments becomes quantitatively acceptable\cite{exp}

Recently, much experimental interest is focused on observation
of probability density function ( in short PDF ) of various
variables in turbulence. It is well-confirmed that the PDF of
velocity differences between a pair of points separated by a
distance clearly deviates from Gaussian and the
deviation becomes more evident for closer pairs. In terms of
Fourier space this means that higher wave number components
of velocity field show larger deviation from Gaussian\cite{katsuyama}.

The multifractal concept and the non-Gaussian PDFs are
consistent representations of the intermittency. Actually,
by combining these two ingredients She and Leveque\cite{empirical}
have derived an analytical formula of multifractal
exponents which fits with experimental values very
nicely. An interesting point in their derivation is that
Navier-Stokes equation is not used explicitly like other
theories on intermittency.

Generally speaking lack of dynamic equation in theory
apparently shows incompleteness of the theory itself,
however, in the present case there is a positive aspect
that the theory's applicability is not limited to
Navier-Stokes turbulence. Actually, almost identical
non-Gaussian PDFs and the multifractal exponents
have been reported in an experiment on thermal
convective turbulence which is governed by
Boussinesq equation\cite{benzi}.

Similar intermittent behavior
can also be found in a drastically simplified numerical
model called GOY model\cite{GOY,Lohse}. It belongs to so-called
shell models as the model discards infinite degrees of
freedom in Navier-Stokes equation and uses only
finite numbers of representative wave number
components called shells. It is described by the
following set of nonlinear differential equations
for complex velocity Fourier components, $u(j,t)$, for
discrete wave numbers $k_j=b^j$,
where $b$ is a positive constant and $j$ is an integer.
\begin{eqnarray}
\frac{d}{dt} u(j,t) &=&
- \nu k_j^2 u(j,t) +f \delta_{j,4}\nonumber\\
&+& i \left\{
a k_j u^*(j+1,t) u^*(j+2,t) \right.\nonumber\\
&&   +bk_{j-1} u^*(j-1,t) u^*(j+1,t)\nonumber\\
&&\left.+c k_{j-2} u^*(j-1,t) u^*(j-2,t)\nonumber
\right\}.
\end{eqnarray}
Here, the first term in right hand side
shows a viscous dissipation effect,
and the second term corresponds to external forcing.
The remaining three terms are nonlinear terms
which are designed to have similar nonlinearity
with Navier-Stokes equation. By these nonlinear
terms this deterministic system shows chaotic
behavior and the resulting statistics agrees with
real fluid intermittency\cite{GOY}.

In this letter, we are going to clarify why such a
simplified model can produce intermittent
behavior quite similar to those in real turbulence
having infinitely many degrees of freedom. Our answer is
simple: The intermittency can be a very general
and universal behavior in a wide class of nonequilibrium
systems not specified to fluid turbulence.
The key point is a fluctuating local directional transfer
in the wave number space, so the number of degrees of
freedom, the details of nonlinearity and the types of
dissipation are not crucial.

In order to validate this scenario
we introduce a new model, the stochastic shell model,
which is even much simpler than GOY model. Instead
of the chaotic nonlinear interactions we introduce discrete
stochastic processes using real variable $u$ as follows;
\begin{eqnarray}
&&u(j,t+1)  = u(j,t) \nonumber\\
&&+ \theta \left[R(j,j-1;t)u(j-1,t) -\mid R(j+1,j;t) \mid u(j,t)\right],
\label{eq:stshell}
\end{eqnarray}
Here, $\theta$ is a non-negative constant less than unity, and
$R(j,j-1;t)$ is a random number which expresses the
momentum transfer from $k_{j-1}$ to $k_j$;
\begin{displaymath}
R(j,j-1;t) = \left \{
\begin{array}{clc}
+1, & \mbox{Prob} & p/2\\
-1, & \mbox{Prob} & p/2\\
0, & \mbox{Prob} & 1-p
\end{array}
\right . ,
\end{displaymath}
It should be noted that
this model can be viewed as a multiplicative process
of  random matrices, thus it does not have any
nonlinearity. Similar linear multiplicative stochastic
process has been analyzed theoretically by Deutsch\cite{matrix}
in which an anomalous scaling relation like the
multifractal scaling is reported.

In the following, we adopt the boundary condition
that $u(N,t)=0$ for $t>0$ where $N$ is the total number of shells.
Compared with the ordinary exponential decay by viscous dissipation
in large wave number regions, this artificial boundary condition
of a sudden cut off may look too rude. However, the boundary
condition at the largest wave number does not affect the
intermediate wave number components because our model
introduces the one-way transport towards the larger wave numbers.
Thus $u(N, t)=0$ is simply for numerical convenience.
As an energy injection effect by an external forcing we add
unifrom random numbers ($\in [-0.5,0.5]$)
 to the 0-th shell at every time step.

Analytical treatment of this stochastic model
can be done by introducing the characteristic
function $Y_j(\rho,t) \equiv \langle \exp [ i \rho u(j,t) ]\rangle$
where $\langle \cal O \rangle$ denotes the averaged value of
the variable $\cal O$ over realizations.
Using this definition we get the following
equation for the characteristic function from Eq.(\ref{eq:stshell}).
\begin{eqnarray}
Y_j (\rho,t+1)& = &\left \{(1-p) Y_j (\rho,t)
+pY_j[(1-\theta)\rho,t] \right \} \nonumber\\
&\times&\left\{(1-p)+pY_{j-1}(\theta\rho,t)\right \}.\nonumber
\end{eqnarray}
We can show analytically that the system quickly converges to a
statistically steady state where all averaged quantities become
independent of time steps. This implies that the ensemble average
and a time average gives the same result independent of initial conditions.
By expanding the characteristic function in terms of $\rho$ up to the
second order we have a rigorous steady state relation for 'energy',
$E_j \equiv \langle \mid u(j,t) \mid^2 /2 \rangle $, as
$ E_j =\theta E_{j-1} / (2-\theta)$.
\begin{figure}
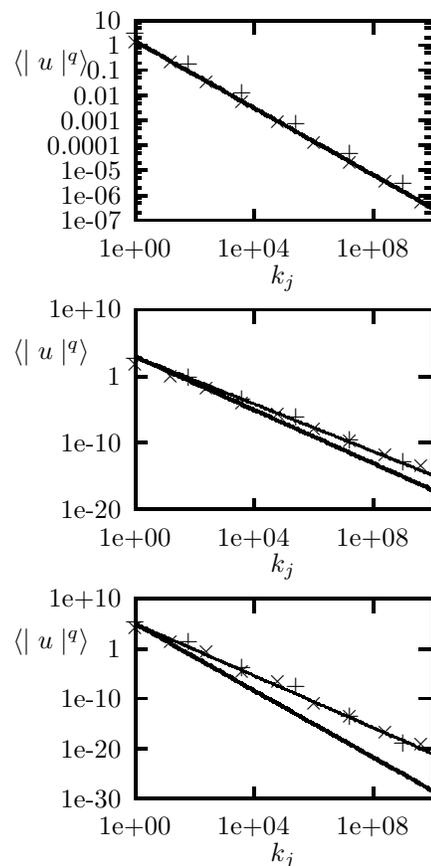

\begin{center}
\setlength{\unitlength}{0.240900pt}
\ifx\plotpoint\undefined\newsavebox{\plotpoint}\fi
\sbox{\plotpoint}{\rule[-0.500pt]{1.000pt}{1.000pt}}%

\caption{Symbols:
Higher moment of $u(j,t)$.(a) $q=2$ (b) $q=6$ (c) $q=10$
( $\times$:b=16,$+$:b=64).
Thick lines: $\zeta_q=q/3$,
thin lines: $\zeta_q= q/9
+ 2\left [ 1-\left ( \frac{2}{3} \right)^{q/3} \right]$.
}
\label{fig:multi}
\end{center}
\end{figure}
The energy spectrum follows the familiar power law form,
$ E_j =k_j^{-\zeta}$, by choosing $\theta$ to be $2/(1+ b^{\zeta})$.
We employ $\zeta=2/3$ so as to satisfy the Kolmogorov spectrum.
The parameter $p$ does not appear in the energy relation and it
turns out that $p$ is irrelevant in the following discussion as far
as $0<p<1$, so we fix its value to be 0.5 hereafter.
Thus, the model has only one free parameter $b$.

To measure the intermittency quantitatively we calculate
higher order moments of $u(j,t)$ and
generalize the definition of the scaling exponent $\zeta$ as
$\langle \mid u(j,t) \mid^q \rangle \propto k_j^{-\zeta_q}$.
Figure \ref{fig:multi} shows  the results for $q= 2$, $6$, and $10$
($N=10, b=16,64$).
Here averages are taken over $2\times 10^5$ steps.
$\zeta_2$ is equal to $2/3$ as expected by analytical treatment.
\begin{figure}
\begin{center}
\setlength{\unitlength}{0.240900pt}
\ifx\plotpoint\undefined\newsavebox{\plotpoint}\fi
\sbox{\plotpoint}{\rule[-0.500pt]{1.000pt}{1.000pt}}%
\begin{picture}(750,450)(0,0)
\font\gnuplot=cmr10 at 10pt
\gnuplot
\sbox{\plotpoint}{\rule[-0.500pt]{1.000pt}{1.000pt}}%
\put(453.0,113.0){\rule[-0.500pt]{1.000pt}{75.643pt}}
\put(220.0,113.0){\rule[-0.500pt]{4.818pt}{1.000pt}}
\put(198,113){\makebox(0,0)[r]{1e-05}}
\put(666.0,113.0){\rule[-0.500pt]{4.818pt}{1.000pt}}
\put(220.0,132.0){\rule[-0.500pt]{2.409pt}{1.000pt}}
\put(676.0,132.0){\rule[-0.500pt]{2.409pt}{1.000pt}}
\put(220.0,157.0){\rule[-0.500pt]{2.409pt}{1.000pt}}
\put(676.0,157.0){\rule[-0.500pt]{2.409pt}{1.000pt}}
\put(220.0,170.0){\rule[-0.500pt]{2.409pt}{1.000pt}}
\put(676.0,170.0){\rule[-0.500pt]{2.409pt}{1.000pt}}
\put(220.0,176.0){\rule[-0.500pt]{4.818pt}{1.000pt}}
\put(198,176){\makebox(0,0)[r]{0.0001}}
\put(666.0,176.0){\rule[-0.500pt]{4.818pt}{1.000pt}}
\put(220.0,195.0){\rule[-0.500pt]{2.409pt}{1.000pt}}
\put(676.0,195.0){\rule[-0.500pt]{2.409pt}{1.000pt}}
\put(220.0,220.0){\rule[-0.500pt]{2.409pt}{1.000pt}}
\put(676.0,220.0){\rule[-0.500pt]{2.409pt}{1.000pt}}
\put(220.0,233.0){\rule[-0.500pt]{2.409pt}{1.000pt}}
\put(676.0,233.0){\rule[-0.500pt]{2.409pt}{1.000pt}}
\put(220.0,239.0){\rule[-0.500pt]{4.818pt}{1.000pt}}
\put(198,239){\makebox(0,0)[r]{0.001}}
\put(666.0,239.0){\rule[-0.500pt]{4.818pt}{1.000pt}}
\put(220.0,258.0){\rule[-0.500pt]{2.409pt}{1.000pt}}
\put(676.0,258.0){\rule[-0.500pt]{2.409pt}{1.000pt}}
\put(220.0,282.0){\rule[-0.500pt]{2.409pt}{1.000pt}}
\put(676.0,282.0){\rule[-0.500pt]{2.409pt}{1.000pt}}
\put(220.0,295.0){\rule[-0.500pt]{2.409pt}{1.000pt}}
\put(676.0,295.0){\rule[-0.500pt]{2.409pt}{1.000pt}}
\put(220.0,301.0){\rule[-0.500pt]{4.818pt}{1.000pt}}
\put(198,301){\makebox(0,0)[r]{0.01}}
\put(666.0,301.0){\rule[-0.500pt]{4.818pt}{1.000pt}}
\put(220.0,320.0){\rule[-0.500pt]{2.409pt}{1.000pt}}
\put(676.0,320.0){\rule[-0.500pt]{2.409pt}{1.000pt}}
\put(220.0,345.0){\rule[-0.500pt]{2.409pt}{1.000pt}}
\put(676.0,345.0){\rule[-0.500pt]{2.409pt}{1.000pt}}
\put(220.0,358.0){\rule[-0.500pt]{2.409pt}{1.000pt}}
\put(676.0,358.0){\rule[-0.500pt]{2.409pt}{1.000pt}}
\put(220.0,364.0){\rule[-0.500pt]{4.818pt}{1.000pt}}
\put(198,364){\makebox(0,0)[r]{0.1}}
\put(666.0,364.0){\rule[-0.500pt]{4.818pt}{1.000pt}}
\put(220.0,383.0){\rule[-0.500pt]{2.409pt}{1.000pt}}
\put(676.0,383.0){\rule[-0.500pt]{2.409pt}{1.000pt}}
\put(220.0,408.0){\rule[-0.500pt]{2.409pt}{1.000pt}}
\put(676.0,408.0){\rule[-0.500pt]{2.409pt}{1.000pt}}
\put(220.0,421.0){\rule[-0.500pt]{2.409pt}{1.000pt}}
\put(676.0,421.0){\rule[-0.500pt]{2.409pt}{1.000pt}}
\put(220.0,427.0){\rule[-0.500pt]{4.818pt}{1.000pt}}
\put(198,427){\makebox(0,0)[r]{1}}
\put(666.0,427.0){\rule[-0.500pt]{4.818pt}{1.000pt}}
\put(220.0,113.0){\rule[-0.500pt]{1.000pt}{4.818pt}}
\put(220,68){\makebox(0,0){-15}}
\put(220.0,407.0){\rule[-0.500pt]{1.000pt}{4.818pt}}
\put(298.0,113.0){\rule[-0.500pt]{1.000pt}{4.818pt}}
\put(298,68){\makebox(0,0){-10}}
\put(298.0,407.0){\rule[-0.500pt]{1.000pt}{4.818pt}}
\put(375.0,113.0){\rule[-0.500pt]{1.000pt}{4.818pt}}
\put(375,68){\makebox(0,0){-5}}
\put(375.0,407.0){\rule[-0.500pt]{1.000pt}{4.818pt}}
\put(453.0,113.0){\rule[-0.500pt]{1.000pt}{4.818pt}}
\put(453,68){\makebox(0,0){0}}
\put(453.0,407.0){\rule[-0.500pt]{1.000pt}{4.818pt}}
\put(531.0,113.0){\rule[-0.500pt]{1.000pt}{4.818pt}}
\put(531,68){\makebox(0,0){5}}
\put(531.0,407.0){\rule[-0.500pt]{1.000pt}{4.818pt}}
\put(608.0,113.0){\rule[-0.500pt]{1.000pt}{4.818pt}}
\put(608,68){\makebox(0,0){10}}
\put(608.0,407.0){\rule[-0.500pt]{1.000pt}{4.818pt}}
\put(686.0,113.0){\rule[-0.500pt]{1.000pt}{4.818pt}}
\put(686,68){\makebox(0,0){15}}
\put(686.0,407.0){\rule[-0.500pt]{1.000pt}{4.818pt}}
\put(220.0,113.0){\rule[-0.500pt]{112.259pt}{1.000pt}}
\put(686.0,113.0){\rule[-0.500pt]{1.000pt}{75.643pt}}
\put(220.0,427.0){\rule[-0.500pt]{112.259pt}{1.000pt}}
\put(111,450){\makebox(0,0){P(u)}}
\put(453,23){\makebox(0,0){u}}
\put(220.0,113.0){\rule[-0.500pt]{1.000pt}{75.643pt}}
\put(387,206){\raisebox{-.8pt}{\makebox(0,0){$\Diamond$}}}
\put(394,238){\raisebox{-.8pt}{\makebox(0,0){$\Diamond$}}}
\put(402,267){\raisebox{-.8pt}{\makebox(0,0){$\Diamond$}}}
\put(410,301){\raisebox{-.8pt}{\makebox(0,0){$\Diamond$}}}
\put(418,330){\raisebox{-.8pt}{\makebox(0,0){$\Diamond$}}}
\put(426,356){\raisebox{-.8pt}{\makebox(0,0){$\Diamond$}}}
\put(433,377){\raisebox{-.8pt}{\makebox(0,0){$\Diamond$}}}
\put(441,393){\raisebox{-.8pt}{\makebox(0,0){$\Diamond$}}}
\put(449,402){\raisebox{-.8pt}{\makebox(0,0){$\Diamond$}}}
\put(457,403){\raisebox{-.8pt}{\makebox(0,0){$\Diamond$}}}
\put(465,395){\raisebox{-.8pt}{\makebox(0,0){$\Diamond$}}}
\put(473,379){\raisebox{-.8pt}{\makebox(0,0){$\Diamond$}}}
\put(480,358){\raisebox{-.8pt}{\makebox(0,0){$\Diamond$}}}
\put(488,334){\raisebox{-.8pt}{\makebox(0,0){$\Diamond$}}}
\put(496,306){\raisebox{-.8pt}{\makebox(0,0){$\Diamond$}}}
\put(504,271){\raisebox{-.8pt}{\makebox(0,0){$\Diamond$}}}
\put(512,236){\raisebox{-.8pt}{\makebox(0,0){$\Diamond$}}}
\put(519,206){\raisebox{-.8pt}{\makebox(0,0){$\Diamond$}}}
\put(527,181){\raisebox{-.8pt}{\makebox(0,0){$\Diamond$}}}
\sbox{\plotpoint}{\rule[-0.175pt]{0.350pt}{0.350pt}}%
\put(382,138){\makebox(0,0){$+$}}
\put(388,191){\makebox(0,0){$+$}}
\put(394,221){\makebox(0,0){$+$}}
\put(400,246){\makebox(0,0){$+$}}
\put(406,282){\makebox(0,0){$+$}}
\put(413,309){\makebox(0,0){$+$}}
\put(419,336){\makebox(0,0){$+$}}
\put(425,357){\makebox(0,0){$+$}}
\put(431,375){\makebox(0,0){$+$}}
\put(437,388){\makebox(0,0){$+$}}
\put(444,397){\makebox(0,0){$+$}}
\put(450,402){\makebox(0,0){$+$}}
\put(456,402){\makebox(0,0){$+$}}
\put(462,397){\makebox(0,0){$+$}}
\put(469,387){\makebox(0,0){$+$}}
\put(475,374){\makebox(0,0){$+$}}
\put(481,358){\makebox(0,0){$+$}}
\put(487,337){\makebox(0,0){$+$}}
\put(493,311){\makebox(0,0){$+$}}
\put(500,284){\makebox(0,0){$+$}}
\put(506,261){\makebox(0,0){$+$}}
\put(512,222){\makebox(0,0){$+$}}
\put(518,210){\makebox(0,0){$+$}}
\put(524,187){\makebox(0,0){$+$}}
\put(531,182){\makebox(0,0){$+$}}
\end{picture}
\setlength{\unitlength}{0.240900pt}
\ifx\plotpoint\undefined\newsavebox{\plotpoint}\fi
\sbox{\plotpoint}{\rule[-0.500pt]{1.000pt}{1.000pt}}%
\begin{picture}(750,450)(0,0)
\font\gnuplot=cmr10 at 10pt
\gnuplot
\sbox{\plotpoint}{\rule[-0.500pt]{1.000pt}{1.000pt}}%
\put(453.0,113.0){\rule[-0.500pt]{1.000pt}{75.643pt}}
\put(220.0,113.0){\rule[-0.500pt]{4.818pt}{1.000pt}}
\put(198,113){\makebox(0,0)[r]{1e-06}}
\put(666.0,113.0){\rule[-0.500pt]{4.818pt}{1.000pt}}
\put(220.0,129.0){\rule[-0.500pt]{2.409pt}{1.000pt}}
\put(676.0,129.0){\rule[-0.500pt]{2.409pt}{1.000pt}}
\put(220.0,150.0){\rule[-0.500pt]{2.409pt}{1.000pt}}
\put(676.0,150.0){\rule[-0.500pt]{2.409pt}{1.000pt}}
\put(220.0,160.0){\rule[-0.500pt]{2.409pt}{1.000pt}}
\put(676.0,160.0){\rule[-0.500pt]{2.409pt}{1.000pt}}
\put(220.0,165.0){\rule[-0.500pt]{4.818pt}{1.000pt}}
\put(198,165){\makebox(0,0)[r]{1e-05}}
\put(666.0,165.0){\rule[-0.500pt]{4.818pt}{1.000pt}}
\put(220.0,181.0){\rule[-0.500pt]{2.409pt}{1.000pt}}
\put(676.0,181.0){\rule[-0.500pt]{2.409pt}{1.000pt}}
\put(220.0,202.0){\rule[-0.500pt]{2.409pt}{1.000pt}}
\put(676.0,202.0){\rule[-0.500pt]{2.409pt}{1.000pt}}
\put(220.0,213.0){\rule[-0.500pt]{2.409pt}{1.000pt}}
\put(676.0,213.0){\rule[-0.500pt]{2.409pt}{1.000pt}}
\put(220.0,218.0){\rule[-0.500pt]{4.818pt}{1.000pt}}
\put(198,218){\makebox(0,0)[r]{0.0001}}
\put(666.0,218.0){\rule[-0.500pt]{4.818pt}{1.000pt}}
\put(220.0,233.0){\rule[-0.500pt]{2.409pt}{1.000pt}}
\put(676.0,233.0){\rule[-0.500pt]{2.409pt}{1.000pt}}
\put(220.0,254.0){\rule[-0.500pt]{2.409pt}{1.000pt}}
\put(676.0,254.0){\rule[-0.500pt]{2.409pt}{1.000pt}}
\put(220.0,265.0){\rule[-0.500pt]{2.409pt}{1.000pt}}
\put(676.0,265.0){\rule[-0.500pt]{2.409pt}{1.000pt}}
\put(220.0,270.0){\rule[-0.500pt]{4.818pt}{1.000pt}}
\put(198,270){\makebox(0,0)[r]{0.001}}
\put(666.0,270.0){\rule[-0.500pt]{4.818pt}{1.000pt}}
\put(220.0,286.0){\rule[-0.500pt]{2.409pt}{1.000pt}}
\put(676.0,286.0){\rule[-0.500pt]{2.409pt}{1.000pt}}
\put(220.0,307.0){\rule[-0.500pt]{2.409pt}{1.000pt}}
\put(676.0,307.0){\rule[-0.500pt]{2.409pt}{1.000pt}}
\put(220.0,317.0){\rule[-0.500pt]{2.409pt}{1.000pt}}
\put(676.0,317.0){\rule[-0.500pt]{2.409pt}{1.000pt}}
\put(220.0,322.0){\rule[-0.500pt]{4.818pt}{1.000pt}}
\put(198,322){\makebox(0,0)[r]{0.01}}
\put(666.0,322.0){\rule[-0.500pt]{4.818pt}{1.000pt}}
\put(220.0,338.0){\rule[-0.500pt]{2.409pt}{1.000pt}}
\put(676.0,338.0){\rule[-0.500pt]{2.409pt}{1.000pt}}
\put(220.0,359.0){\rule[-0.500pt]{2.409pt}{1.000pt}}
\put(676.0,359.0){\rule[-0.500pt]{2.409pt}{1.000pt}}
\put(220.0,370.0){\rule[-0.500pt]{2.409pt}{1.000pt}}
\put(676.0,370.0){\rule[-0.500pt]{2.409pt}{1.000pt}}
\put(220.0,375.0){\rule[-0.500pt]{4.818pt}{1.000pt}}
\put(198,375){\makebox(0,0)[r]{0.1}}
\put(666.0,375.0){\rule[-0.500pt]{4.818pt}{1.000pt}}
\put(220.0,390.0){\rule[-0.500pt]{2.409pt}{1.000pt}}
\put(676.0,390.0){\rule[-0.500pt]{2.409pt}{1.000pt}}
\put(220.0,411.0){\rule[-0.500pt]{2.409pt}{1.000pt}}
\put(676.0,411.0){\rule[-0.500pt]{2.409pt}{1.000pt}}
\put(220.0,422.0){\rule[-0.500pt]{2.409pt}{1.000pt}}
\put(676.0,422.0){\rule[-0.500pt]{2.409pt}{1.000pt}}
\put(220.0,427.0){\rule[-0.500pt]{4.818pt}{1.000pt}}
\put(198,427){\makebox(0,0)[r]{1}}
\put(666.0,427.0){\rule[-0.500pt]{4.818pt}{1.000pt}}
\put(220.0,113.0){\rule[-0.500pt]{1.000pt}{4.818pt}}
\put(220,68){\makebox(0,0){-15}}
\put(220.0,407.0){\rule[-0.500pt]{1.000pt}{4.818pt}}
\put(298.0,113.0){\rule[-0.500pt]{1.000pt}{4.818pt}}
\put(298,68){\makebox(0,0){-10}}
\put(298.0,407.0){\rule[-0.500pt]{1.000pt}{4.818pt}}
\put(375.0,113.0){\rule[-0.500pt]{1.000pt}{4.818pt}}
\put(375,68){\makebox(0,0){-5}}
\put(375.0,407.0){\rule[-0.500pt]{1.000pt}{4.818pt}}
\put(453.0,113.0){\rule[-0.500pt]{1.000pt}{4.818pt}}
\put(453,68){\makebox(0,0){0}}
\put(453.0,407.0){\rule[-0.500pt]{1.000pt}{4.818pt}}
\put(531.0,113.0){\rule[-0.500pt]{1.000pt}{4.818pt}}
\put(531,68){\makebox(0,0){5}}
\put(531.0,407.0){\rule[-0.500pt]{1.000pt}{4.818pt}}
\put(608.0,113.0){\rule[-0.500pt]{1.000pt}{4.818pt}}
\put(608,68){\makebox(0,0){10}}
\put(608.0,407.0){\rule[-0.500pt]{1.000pt}{4.818pt}}
\put(686.0,113.0){\rule[-0.500pt]{1.000pt}{4.818pt}}
\put(686,68){\makebox(0,0){15}}
\put(686.0,407.0){\rule[-0.500pt]{1.000pt}{4.818pt}}
\put(220.0,113.0){\rule[-0.500pt]{112.259pt}{1.000pt}}
\put(686.0,113.0){\rule[-0.500pt]{1.000pt}{75.643pt}}
\put(220.0,427.0){\rule[-0.500pt]{112.259pt}{1.000pt}}
\put(111,450){\makebox(0,0){P(u)}}
\put(453,23){\makebox(0,0){u}}
\put(220.0,113.0){\rule[-0.500pt]{1.000pt}{75.643pt}}
\put(319,147){\raisebox{-.8pt}{\makebox(0,0){$\Diamond$}}}
\put(330,147){\raisebox{-.8pt}{\makebox(0,0){$\Diamond$}}}
\put(340,163){\raisebox{-.8pt}{\makebox(0,0){$\Diamond$}}}
\put(351,194){\raisebox{-.8pt}{\makebox(0,0){$\Diamond$}}}
\put(362,205){\raisebox{-.8pt}{\makebox(0,0){$\Diamond$}}}
\put(373,234){\raisebox{-.8pt}{\makebox(0,0){$\Diamond$}}}
\put(383,261){\raisebox{-.8pt}{\makebox(0,0){$\Diamond$}}}
\put(394,287){\raisebox{-.8pt}{\makebox(0,0){$\Diamond$}}}
\put(405,313){\raisebox{-.8pt}{\makebox(0,0){$\Diamond$}}}
\put(415,338){\raisebox{-.8pt}{\makebox(0,0){$\Diamond$}}}
\put(426,364){\raisebox{-.8pt}{\makebox(0,0){$\Diamond$}}}
\put(437,388){\raisebox{-.8pt}{\makebox(0,0){$\Diamond$}}}
\put(448,407){\raisebox{-.8pt}{\makebox(0,0){$\Diamond$}}}
\put(458,409){\raisebox{-.8pt}{\makebox(0,0){$\Diamond$}}}
\put(469,392){\raisebox{-.8pt}{\makebox(0,0){$\Diamond$}}}
\put(480,368){\raisebox{-.8pt}{\makebox(0,0){$\Diamond$}}}
\put(491,343){\raisebox{-.8pt}{\makebox(0,0){$\Diamond$}}}
\put(501,317){\raisebox{-.8pt}{\makebox(0,0){$\Diamond$}}}
\put(512,290){\raisebox{-.8pt}{\makebox(0,0){$\Diamond$}}}
\put(523,265){\raisebox{-.8pt}{\makebox(0,0){$\Diamond$}}}
\put(533,240){\raisebox{-.8pt}{\makebox(0,0){$\Diamond$}}}
\put(544,214){\raisebox{-.8pt}{\makebox(0,0){$\Diamond$}}}
\put(555,183){\raisebox{-.8pt}{\makebox(0,0){$\Diamond$}}}
\put(566,163){\raisebox{-.8pt}{\makebox(0,0){$\Diamond$}}}
\put(576,163){\raisebox{-.8pt}{\makebox(0,0){$\Diamond$}}}
\put(587,147){\raisebox{-.8pt}{\makebox(0,0){$\Diamond$}}}
\sbox{\plotpoint}{\rule[-0.175pt]{0.350pt}{0.350pt}}%
\put(350,211){\makebox(0,0){$+$}}
\put(363,241){\makebox(0,0){$+$}}
\put(369,236){\makebox(0,0){$+$}}
\put(375,258){\makebox(0,0){$+$}}
\put(382,273){\makebox(0,0){$+$}}
\put(388,286){\makebox(0,0){$+$}}
\put(394,294){\makebox(0,0){$+$}}
\put(400,312){\makebox(0,0){$+$}}
\put(406,319){\makebox(0,0){$+$}}
\put(413,333){\makebox(0,0){$+$}}
\put(419,348){\makebox(0,0){$+$}}
\put(425,362){\makebox(0,0){$+$}}
\put(431,376){\makebox(0,0){$+$}}
\put(437,390){\makebox(0,0){$+$}}
\put(444,403){\makebox(0,0){$+$}}
\put(450,410){\makebox(0,0){$+$}}
\put(456,410){\makebox(0,0){$+$}}
\put(462,403){\makebox(0,0){$+$}}
\put(469,390){\makebox(0,0){$+$}}
\put(475,376){\makebox(0,0){$+$}}
\put(481,361){\makebox(0,0){$+$}}
\put(487,347){\makebox(0,0){$+$}}
\put(493,334){\makebox(0,0){$+$}}
\put(500,322){\makebox(0,0){$+$}}
\put(506,313){\makebox(0,0){$+$}}
\put(512,294){\makebox(0,0){$+$}}
\put(518,277){\makebox(0,0){$+$}}
\put(524,268){\makebox(0,0){$+$}}
\put(531,255){\makebox(0,0){$+$}}
\put(537,241){\makebox(0,0){$+$}}
\put(543,230){\makebox(0,0){$+$}}
\put(562,186){\makebox(0,0){$+$}}
\end{picture}
\setlength{\unitlength}{0.240900pt}
\ifx\plotpoint\undefined\newsavebox{\plotpoint}\fi
\sbox{\plotpoint}{\rule[-0.500pt]{1.000pt}{1.000pt}}%
\begin{picture}(750,450)(0,0)
\font\gnuplot=cmr10 at 10pt
\gnuplot
\sbox{\plotpoint}{\rule[-0.500pt]{1.000pt}{1.000pt}}%
\put(453.0,113.0){\rule[-0.500pt]{1.000pt}{75.643pt}}
\put(220.0,113.0){\rule[-0.500pt]{4.818pt}{1.000pt}}
\put(198,113){\makebox(0,0)[r]{1e-06}}
\put(666.0,113.0){\rule[-0.500pt]{4.818pt}{1.000pt}}
\put(220.0,129.0){\rule[-0.500pt]{2.409pt}{1.000pt}}
\put(676.0,129.0){\rule[-0.500pt]{2.409pt}{1.000pt}}
\put(220.0,150.0){\rule[-0.500pt]{2.409pt}{1.000pt}}
\put(676.0,150.0){\rule[-0.500pt]{2.409pt}{1.000pt}}
\put(220.0,160.0){\rule[-0.500pt]{2.409pt}{1.000pt}}
\put(676.0,160.0){\rule[-0.500pt]{2.409pt}{1.000pt}}
\put(220.0,165.0){\rule[-0.500pt]{4.818pt}{1.000pt}}
\put(198,165){\makebox(0,0)[r]{1e-05}}
\put(666.0,165.0){\rule[-0.500pt]{4.818pt}{1.000pt}}
\put(220.0,181.0){\rule[-0.500pt]{2.409pt}{1.000pt}}
\put(676.0,181.0){\rule[-0.500pt]{2.409pt}{1.000pt}}
\put(220.0,202.0){\rule[-0.500pt]{2.409pt}{1.000pt}}
\put(676.0,202.0){\rule[-0.500pt]{2.409pt}{1.000pt}}
\put(220.0,213.0){\rule[-0.500pt]{2.409pt}{1.000pt}}
\put(676.0,213.0){\rule[-0.500pt]{2.409pt}{1.000pt}}
\put(220.0,218.0){\rule[-0.500pt]{4.818pt}{1.000pt}}
\put(198,218){\makebox(0,0)[r]{0.0001}}
\put(666.0,218.0){\rule[-0.500pt]{4.818pt}{1.000pt}}
\put(220.0,233.0){\rule[-0.500pt]{2.409pt}{1.000pt}}
\put(676.0,233.0){\rule[-0.500pt]{2.409pt}{1.000pt}}
\put(220.0,254.0){\rule[-0.500pt]{2.409pt}{1.000pt}}
\put(676.0,254.0){\rule[-0.500pt]{2.409pt}{1.000pt}}
\put(220.0,265.0){\rule[-0.500pt]{2.409pt}{1.000pt}}
\put(676.0,265.0){\rule[-0.500pt]{2.409pt}{1.000pt}}
\put(220.0,270.0){\rule[-0.500pt]{4.818pt}{1.000pt}}
\put(198,270){\makebox(0,0)[r]{0.001}}
\put(666.0,270.0){\rule[-0.500pt]{4.818pt}{1.000pt}}
\put(220.0,286.0){\rule[-0.500pt]{2.409pt}{1.000pt}}
\put(676.0,286.0){\rule[-0.500pt]{2.409pt}{1.000pt}}
\put(220.0,307.0){\rule[-0.500pt]{2.409pt}{1.000pt}}
\put(676.0,307.0){\rule[-0.500pt]{2.409pt}{1.000pt}}
\put(220.0,317.0){\rule[-0.500pt]{2.409pt}{1.000pt}}
\put(676.0,317.0){\rule[-0.500pt]{2.409pt}{1.000pt}}
\put(220.0,322.0){\rule[-0.500pt]{4.818pt}{1.000pt}}
\put(198,322){\makebox(0,0)[r]{0.01}}
\put(666.0,322.0){\rule[-0.500pt]{4.818pt}{1.000pt}}
\put(220.0,338.0){\rule[-0.500pt]{2.409pt}{1.000pt}}
\put(676.0,338.0){\rule[-0.500pt]{2.409pt}{1.000pt}}
\put(220.0,359.0){\rule[-0.500pt]{2.409pt}{1.000pt}}
\put(676.0,359.0){\rule[-0.500pt]{2.409pt}{1.000pt}}
\put(220.0,370.0){\rule[-0.500pt]{2.409pt}{1.000pt}}
\put(676.0,370.0){\rule[-0.500pt]{2.409pt}{1.000pt}}
\put(220.0,375.0){\rule[-0.500pt]{4.818pt}{1.000pt}}
\put(198,375){\makebox(0,0)[r]{0.1}}
\put(666.0,375.0){\rule[-0.500pt]{4.818pt}{1.000pt}}
\put(220.0,390.0){\rule[-0.500pt]{2.409pt}{1.000pt}}
\put(676.0,390.0){\rule[-0.500pt]{2.409pt}{1.000pt}}
\put(220.0,411.0){\rule[-0.500pt]{2.409pt}{1.000pt}}
\put(676.0,411.0){\rule[-0.500pt]{2.409pt}{1.000pt}}
\put(220.0,422.0){\rule[-0.500pt]{2.409pt}{1.000pt}}
\put(676.0,422.0){\rule[-0.500pt]{2.409pt}{1.000pt}}
\put(220.0,427.0){\rule[-0.500pt]{4.818pt}{1.000pt}}
\put(198,427){\makebox(0,0)[r]{1}}
\put(666.0,427.0){\rule[-0.500pt]{4.818pt}{1.000pt}}
\put(220.0,113.0){\rule[-0.500pt]{1.000pt}{4.818pt}}
\put(220,68){\makebox(0,0){-15}}
\put(220.0,407.0){\rule[-0.500pt]{1.000pt}{4.818pt}}
\put(298.0,113.0){\rule[-0.500pt]{1.000pt}{4.818pt}}
\put(298,68){\makebox(0,0){-10}}
\put(298.0,407.0){\rule[-0.500pt]{1.000pt}{4.818pt}}
\put(375.0,113.0){\rule[-0.500pt]{1.000pt}{4.818pt}}
\put(375,68){\makebox(0,0){-5}}
\put(375.0,407.0){\rule[-0.500pt]{1.000pt}{4.818pt}}
\put(453.0,113.0){\rule[-0.500pt]{1.000pt}{4.818pt}}
\put(453,68){\makebox(0,0){0}}
\put(453.0,407.0){\rule[-0.500pt]{1.000pt}{4.818pt}}
\put(531.0,113.0){\rule[-0.500pt]{1.000pt}{4.818pt}}
\put(531,68){\makebox(0,0){5}}
\put(531.0,407.0){\rule[-0.500pt]{1.000pt}{4.818pt}}
\put(608.0,113.0){\rule[-0.500pt]{1.000pt}{4.818pt}}
\put(608,68){\makebox(0,0){10}}
\put(608.0,407.0){\rule[-0.500pt]{1.000pt}{4.818pt}}
\put(686.0,113.0){\rule[-0.500pt]{1.000pt}{4.818pt}}
\put(686,68){\makebox(0,0){15}}
\put(686.0,407.0){\rule[-0.500pt]{1.000pt}{4.818pt}}
\put(220.0,113.0){\rule[-0.500pt]{112.259pt}{1.000pt}}
\put(686.0,113.0){\rule[-0.500pt]{1.000pt}{75.643pt}}
\put(220.0,427.0){\rule[-0.500pt]{112.259pt}{1.000pt}}
\put(111,450){\makebox(0,0){P(u)}}
\put(453,23){\makebox(0,0){u}}
\put(220.0,113.0){\rule[-0.500pt]{1.000pt}{75.643pt}}
\put(250,120){\raisebox{-.8pt}{\makebox(0,0){$\Diamond$}}}
\put(275,136){\raisebox{-.8pt}{\makebox(0,0){$\Diamond$}}}
\put(287,165){\raisebox{-.8pt}{\makebox(0,0){$\Diamond$}}}
\put(300,173){\raisebox{-.8pt}{\makebox(0,0){$\Diamond$}}}
\put(312,200){\raisebox{-.8pt}{\makebox(0,0){$\Diamond$}}}
\put(324,202){\raisebox{-.8pt}{\makebox(0,0){$\Diamond$}}}
\put(336,216){\raisebox{-.8pt}{\makebox(0,0){$\Diamond$}}}
\put(349,236){\raisebox{-.8pt}{\makebox(0,0){$\Diamond$}}}
\put(361,253){\raisebox{-.8pt}{\makebox(0,0){$\Diamond$}}}
\put(373,268){\raisebox{-.8pt}{\makebox(0,0){$\Diamond$}}}
\put(385,286){\raisebox{-.8pt}{\makebox(0,0){$\Diamond$}}}
\put(398,305){\raisebox{-.8pt}{\makebox(0,0){$\Diamond$}}}
\put(410,324){\raisebox{-.8pt}{\makebox(0,0){$\Diamond$}}}
\put(422,347){\raisebox{-.8pt}{\makebox(0,0){$\Diamond$}}}
\put(435,373){\raisebox{-.8pt}{\makebox(0,0){$\Diamond$}}}
\put(447,406){\raisebox{-.8pt}{\makebox(0,0){$\Diamond$}}}
\put(459,413){\raisebox{-.8pt}{\makebox(0,0){$\Diamond$}}}
\put(471,379){\raisebox{-.8pt}{\makebox(0,0){$\Diamond$}}}
\put(484,353){\raisebox{-.8pt}{\makebox(0,0){$\Diamond$}}}
\put(496,330){\raisebox{-.8pt}{\makebox(0,0){$\Diamond$}}}
\put(508,309){\raisebox{-.8pt}{\makebox(0,0){$\Diamond$}}}
\put(521,290){\raisebox{-.8pt}{\makebox(0,0){$\Diamond$}}}
\put(533,274){\raisebox{-.8pt}{\makebox(0,0){$\Diamond$}}}
\put(545,256){\raisebox{-.8pt}{\makebox(0,0){$\Diamond$}}}
\put(557,238){\raisebox{-.8pt}{\makebox(0,0){$\Diamond$}}}
\put(570,224){\raisebox{-.8pt}{\makebox(0,0){$\Diamond$}}}
\put(582,204){\raisebox{-.8pt}{\makebox(0,0){$\Diamond$}}}
\put(594,191){\raisebox{-.8pt}{\makebox(0,0){$\Diamond$}}}
\put(606,182){\raisebox{-.8pt}{\makebox(0,0){$\Diamond$}}}
\put(619,152){\raisebox{-.8pt}{\makebox(0,0){$\Diamond$}}}
\put(631,157){\raisebox{-.8pt}{\makebox(0,0){$\Diamond$}}}
\put(643,145){\raisebox{-.8pt}{\makebox(0,0){$\Diamond$}}}
\put(656,120){\raisebox{-.8pt}{\makebox(0,0){$\Diamond$}}}
\put(668,120){\raisebox{-.8pt}{\makebox(0,0){$\Diamond$}}}
\sbox{\plotpoint}{\rule[-0.175pt]{0.350pt}{0.350pt}}%
\put(332,202){\makebox(0,0){$+$}}
\put(344,223){\makebox(0,0){$+$}}
\put(350,218){\makebox(0,0){$+$}}
\put(357,230){\makebox(0,0){$+$}}
\put(363,246){\makebox(0,0){$+$}}
\put(369,261){\makebox(0,0){$+$}}
\put(375,271){\makebox(0,0){$+$}}
\put(382,271){\makebox(0,0){$+$}}
\put(388,297){\makebox(0,0){$+$}}
\put(394,303){\makebox(0,0){$+$}}
\put(400,312){\makebox(0,0){$+$}}
\put(406,320){\makebox(0,0){$+$}}
\put(413,332){\makebox(0,0){$+$}}
\put(419,346){\makebox(0,0){$+$}}
\put(425,358){\makebox(0,0){$+$}}
\put(431,370){\makebox(0,0){$+$}}
\put(437,385){\makebox(0,0){$+$}}
\put(444,401){\makebox(0,0){$+$}}
\put(450,415){\makebox(0,0){$+$}}
\put(456,414){\makebox(0,0){$+$}}
\put(462,401){\makebox(0,0){$+$}}
\put(469,386){\makebox(0,0){$+$}}
\put(475,371){\makebox(0,0){$+$}}
\put(481,356){\makebox(0,0){$+$}}
\put(487,343){\makebox(0,0){$+$}}
\put(493,332){\makebox(0,0){$+$}}
\put(500,322){\makebox(0,0){$+$}}
\put(506,311){\makebox(0,0){$+$}}
\put(512,300){\makebox(0,0){$+$}}
\put(518,290){\makebox(0,0){$+$}}
\put(524,280){\makebox(0,0){$+$}}
\put(531,270){\makebox(0,0){$+$}}
\put(537,260){\makebox(0,0){$+$}}
\put(543,263){\makebox(0,0){$+$}}
\put(549,251){\makebox(0,0){$+$}}
\put(556,251){\makebox(0,0){$+$}}
\put(562,246){\makebox(0,0){$+$}}
\put(568,236){\makebox(0,0){$+$}}
\put(574,223){\makebox(0,0){$+$}}
\put(587,211){\makebox(0,0){$+$}}
\put(593,218){\makebox(0,0){$+$}}
\put(599,186){\makebox(0,0){$+$}}
\put(605,202){\makebox(0,0){$+$}}
\put(611,211){\makebox(0,0){$+$}}
\put(624,202){\makebox(0,0){$+$}}
\put(630,186){\makebox(0,0){$+$}}
\put(655,186){\makebox(0,0){$+$}}
\end{picture}
\end{center}
\caption{Comparison of PDF between stochastic shell model ($+$)
and experiment ($\diamond$).
All data are normalized so as to have variance of unity
for comparison.
Top: Experiment: 0.15kHz, model: $j=0$.
Middle: Experiment: 3kHz, model: $j=1$.
Bottom: Experiment: 10kHz, model: $j=2$.
}
\label{fig:PDFs}
\end{figure}
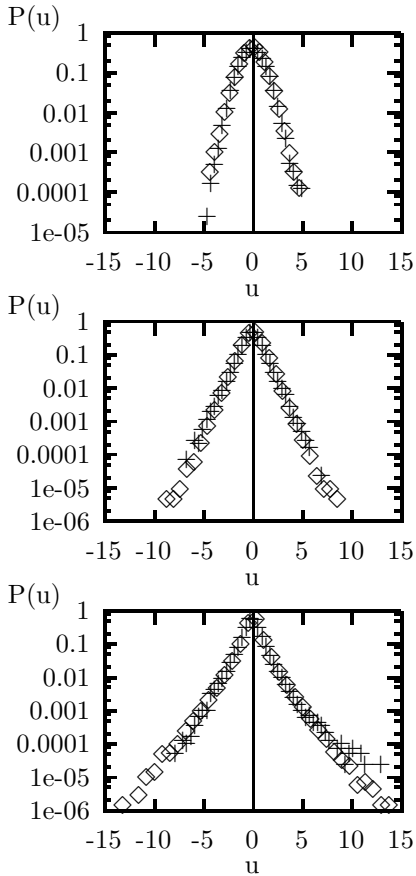
On the other hand,
values of $\zeta_q$ for $q = 6$ and $10$ clearly deviate from
Kolmogorov's original scaling exponents, $q/3$.
We can conclude that our model exhibits an anomalous scaling\cite{unique}.
In order to compare the exponents
with experimental values,
we also plot the empirical multifractal relation,
$\zeta_q= q/9 + 2\left [ 1-\left ( \frac{2}{3} \right)^{q/3} \right]$,
which is known to be a good approximation to the experimental values
\cite{empirical}.
Our data agree with this empirical values in the whole range, meaning
that our model can reproduce the multifractal scaling quantitatively.

Next, we check our model whether the wave number dependence of
PDFs observed in experiments can also be reproduced correctly or not.
Figure \ref{fig:PDFs} shows a
comparison between experimentally observed PDFs
and those of our model ($b=64, N=5$). The experimental turbulence
data are PDFs of frequency-band-pass-filtered velocity signals
obtained by one of the authors TK\cite{katsuyama}. In the case of
our numerical model we plotted 
 $u(j,t)$ because
the imaginary part follows the same statistics.
Data are sampled every two steps over total $2 \times 10^5$ steps.
The experimental PDFs show a tendency that the deviation
from Gaussian becomes more evident for larger wave numbers.
The numerical data fit them very nicely in the whole range in
each case. Namely, our model also reproduces the other aspect
of intermittency, the non-Gaussian PDFs\cite{RAD}.

In high wave number limit
it is known that both GOY model and
Navier-Stokes turbulence
have almost identical power laws in PDF $P(u)$ of velocity $u$,
i.e., $P(u) \sim u^{-\beta}$ with $\beta \simeq 1.6$\cite{KYO}.
Thus, the appearance of power tails in PDFs at very high wavenumber
may also be an important nature of fluid intermittency.
\begin{figure}
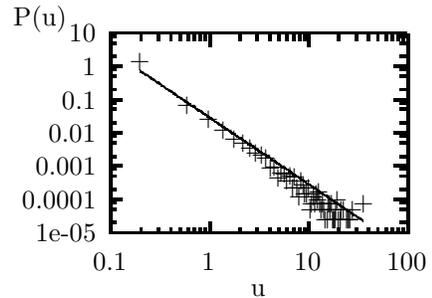

\begin{center}
\setlength{\unitlength}{0.240900pt}
\ifx\plotpoint\undefined\newsavebox{\plotpoint}\fi
\sbox{\plotpoint}{\rule[-0.500pt]{1.000pt}{1.000pt}}%

\end{center}
\caption{Power law PDF obtained by the stochastic shell model
($j=40$). The straight line indicates $P(u) \sim u^{-2}$}
\label{fig:powerPDF}
\end{figure}
Figure \ref{fig:powerPDF} shows
PDF of our model in very high wave number region ($j=40$, $N=50, b=16$).
Data are sampled every two steps over total $2 \times 10^5$ steps.
The PDF shows a power law with its exponent
about $-2$. Taking into account the errors accompanied with
the estimation of such delicate exponent, this value is not far
from the values of fluid turbulence, $-1.6$. We judge that our
model is also consistent on this power law behavior\cite{analytic}.
Remembering that our model is linear and has no
viscosity, the above results suggests that the intermittency
may not have any direct relationship with nonlinearity and
viscosity. This may sounds quite contrary to the common sense, and
a big question naturally arises, that is, what is really responsible for
the intermittency?

To elucidate minimal factors for the realization of intermittency
we still modify the model. Instead of real variables we consider
complex variables for $u(j,t)$ in our stochastic shell model.
The results are almost identical with what
we have reported.
However, when we fix the sign of $R(j,j-1;t)$,
independent of whether $u$ is real or complex,
we get considerably different state.
We have a statistically steady result
and the distributions deviate more from Gaussian for larger
wave number components as in the case with sign randomization,
but the energy spectrum is modified to a
non-power law and the multifractal scalings are lost
completely. This difference is expected to be caused by the
strong correlations among velocity components in each shell.
This result demonstrates that
transport with sign (or phase for complex $u$)
randomization is essential for the
intermittency in real turbulence.

Concluding the paper, we have introduced a stochastic shell model to
describe cascading transports in wave number space,
which can be viewed as a kind of linear multiplicative
stochastic process.
The model with sign randomization reproduces the intermittent behavior
correctly while its constant sign
version looses all such behavior. This fact
implies that the origin of intermittency is in the cascading process
with sign (or phase for complex $u$)
. Nonlinearity and viscosity does not appear explicitly in
our discussion, therefore,
these effects may give only indirect contribution to
the intermittency.
Non-linearity in real turbulence is important as
a origin of randomness.
However, once randomness is introduced,
it is not important whether randomness originates in
non-linearity or white noise when
we consider stochastic property like
intermittency and anomalous scaling.
Further investigation of the stochastic shell model
will contribute on the problem of universality of intermittency.

HT thanks Dr. Misako Takayasu for helpful discussions.
YHT acknowledge Prof. M. Yamada for stimulating discussions and
people who developed Linux without which we cannot perform this research.

\end{multicols}

\begin{references}
\bibitem{unique}
One should notice that values of $\zeta_q$ is almost
independent of $b$, thus
there are no free parameters on the decision of
values of $\zeta_q$.
\end{references}
\end{document}